\documentclass[a4paper,11pt]{article}
\usepackage{latexsym}
\usepackage[utf8]{inputenc}
\usepackage[activeacute,spanish]{babel}
\usepackage{lineno}
\usepackage{authblk}

\usepackage[
    style=phys,
    articletitle=true,biblabel=brackets,
    chaptertitle=false,pageranges=false
]{biblatex}
\bibliography{articulo}
\RequirePackage{graphicx}
\graphicspath{ {./imagenes/} }
\RequirePackage{booktabs}

\usepackage[colorlinks,linkcolor={blue},citecolor={blue},urlcolor={red}]{hyperref}
\hypersetup{urlcolor=blue, colorlinks=true} 

\usepackage{geometry} 
\title{CORRELACIÓN DE MATERIAL PARTICULADO $PM_{2.5}$ EN FUNCIÓN A LA HUMEDAD RELATIVA - PRECIPITACIÓN USANDO Python, EN LA CIUDAD DEL CUSCO}
\author[1]{Bruce Warthon}
\author[1]{Iván Miranda}
\author[1]{Iván R. Quispe}
\author[1]{Rafael Ponce}
\author[1]{Victor F. Ramos}
\author[1]{Ariatna Zamalloa}
\author[1]{Julio Warthon}
\author[1]{Ruben Tupayachi}
    \author[1]{Miluska Zamalloa}
\affil[1]{Universidad Nacional de San Antonio Abad del Cusco}
\affil[1]{160531@unsaac.edu.pe}
\affil[2]{150353@unsaac.edu.pe}
\affil[3]{161729@unsaac.edu.pe}
\affil[4]{161316@unsaac.edu.pe}
\affil[5]{160855@unsaac.edu.pe}
\affil[6]{182773@unsaac.edu.pe}
\affil[7]{julio.warthon@unsaac.edu.pe}
\affil[8]{ruben.tupayachi@unsaac.edu.pe}
\affil[9]{miluska.zamalloa@unsaac.edu.pe}

\usepackage{float}
\usepackage{paralist}
\usepackage{multirow}
\usepackage{csquotes}

\addto\captionsspanish{
}

\begin{document}
\maketitle

\begin{abstract} \noindent 
El presente estudio utiliza datos correspondientes al promedio de concentración mensual de material particulado 2.5 ($PM_{2.5}$), humedad relativa y precipitación durante el periodo 2016-2018 recolectados por la estación High Vol del Centro de Energía y Atmósfera - UNSAAC, y el centro meteorologico SENAHMI (Kayra-UNSAAC). Para estimar la relación estadística entre dos variables cuantitativas, se utiliza el coeficiente de correlación de Pearson y Spearman, los cuales fueron obtenidos mediante el lenguaje de programación Python. El valor obtenido de t se compara con el valor obtenido de la tabla t de Student para un cierto nivel de significancia $(\alpha = 0,05)$. Se establecen valores de correlación positiva y negativa en función de los coeficientes obtenidos, y se interpretan los valores de correlación de Pearson y Spearman para determinar si los datos están o no relacionados entre si. Se comprueba finalmente nuestra hipótesis, estableciendo una alta correlación negativa entre valores de los promedios mensuales del material particulado en función a la humedad relativa y la precipitación.\\
\textbf{Palabras clave}: Material particulado 2.5 ($PM_{2.5}$), Humedad relativa
, Precipitación, Coeficiente de correlación de Pearson, Coeficiente de correlación de Spearman, Python.
\end{abstract}

\renewcommand{\abstractname}{Abstract}

\begin{abstract} \noindent
This study uses data corresponding to the monthly average concentration of particulate matter 2.5 ($PM_{2.5}$), relative humidity, and precipitation during the period 2016-2018 collected by the High Vol station of the Energy and Atmosphere Center - UNSAAC and the meteorological center SENAHMI (Kayra-UNSAAC). In order to estimate the statistical relationship between two quantitative variables, the Pearson and Spearman correlation coefficients are used, which were obtained using the Python programming language. The obtained t-value is compared with the value obtained from the Student's t-table for a certain level of significance $(\alpha = 0.05)$. Positive and negative correlation values are established based on the obtained coefficients, and the Pearson and Spearman correlation values are interpreted to determine whether the data are related or not. Finally, our hypothesis is verified by establishing a high negative correlation between monthly average values of particulate matter based on relative humidity and precipitation.
\end{abstract}

\section{Introducción}
En las ultimas décadas la ciudad de Cusco viene experimentado un deterioro en la calidad del aire atmosférica causado principalmente por el material particulado, los valores de las concentraciones de los contaminantes atmosféricos han alcanzado valores que conllevan
riesgo para la salud de las personas. Según estudios de la Organización Mundial de la Salud (OMS), la exposición prolongada a $PM_{2.5}$ puede tener efectos negativos en la salud, como la disminución de la función pulmonar, el aumento del riesgo de enfermedades cardiovasculares y la mortalidad prematura ~\cite{guias}. Las partículas finas como $PM_{2.5}$ pueden penetrar profundamente hasta los alveolos pulmonares de los humanos conllevando a riesgos de salud pública. Además, las $PM_{2.5}$ también tienen un impacto significativo en el medio ambiente, contribuyendo al cambio climático y la acidificación del suelo y del agua~\cite{Epa_2016}.

Los altos niveles de contaminación atmosférica en el Cusco se deben a la unión de varios factores como el incremento del parque automotor en las últimas décadas, sobre todo con vehículos de segundo uso~\cite{mora2018calidad}. Los valores proporcionados por la SUNARP y el ENEI muestran que en 1989 habían 11,806 vehículos
en circulación, en el 2001 el número fue de 33,316 unidades motorizadas, el crecimiento
fue de 282 \% en el lapso de 12 años, en el 2010 fue de 85,560 vehículos~\cite{mora2018calidad}.

La concentración de los contaminantes en la atmósfera esta afectada por variables meteorológicas como, la humedad relativa y precipitación. Por otro lado, la dispersión de estos contaminantes es influenciada por la precipitación y la humeada relativa, así como la estabilidad que predomina en la atmósfera. La hipótesis de la investigación es una posible correlación de la concentración de $PM_{2.5}$ en el aire y la humedad relativa conjuntamente con la precipitación. En ciudades de altura por encima de los 3000 metros sobre el nivel del mar (msnm) la contaminación del aire puede ser una preocupación particular debido a factores como la menor densidad del aire, la topografía y las condiciones climáticas. A nivel regional se ha puesto atención al problema de la contaminación atmosférica por lo tanto a través de una red de monitoreo automatizada conocida como Sistema de monitoreo Shelter del Centro de Investigación de Energía y Atmosfera - UNSAAC en la ciudad del Cusco. Esta red de monitoreo opero desde 2016 al 2018 con el objetivo de muestrear las concentraciones del material particulado ($PM_{2.5}$) ubicadas en las periferias de las de la ciudad el cusco. Referente a la condiciones climáticas en la ciudad del cusco la cual está ubicada en una plataforma andina a una altitud de 3,340 m.s.n.m y está a lo largo  de  una cuenca de 31 km y está rodeado por colinas hacia el norte y oeste y abierta al sur y de esta forma está condicionada para la formación de un bolsón de aire seco donde la dirección predominante de los vientos es de sur a norte~\cite{carac_clima}. Por lo tanto, en este estudio, se analizarán los datos de $PM_{2.5}$, precipitación y humedad relativa recopilados durante un período de dos años para determinar la correlación entre estos factores en la ciudad del Cusco a mas de 3000msnm. Los resultados de este estudio proporcionan información valiosa sobre los factores que influyen en la contaminación del aire en ciudades de alta altitud y contribuyen a la toma de decisiones informadas sobre políticas y medidas de control de la contaminación del aire en estas áreas.

Los estudios anteriores han encontrado que la precipitación puede reducir la concentración de $PM_{2.5}$ en el aire. Por ejemplo, en un estudio realizado en la ciudad de Wuhan, China,~\cite{Zhang_2020}. Encontraron que la precipitación tenía un efecto significativo en la reducción de la concentración de $PM_{2.5}$ en el aire. Adicionalmente Liu~\cite{Liu_2017}, encontró una relación inversa entre la concentración de $PM_{2.5}$ y la precipitación, lo que indica que la lluvia puede eliminar las partículas del aire en un estudio realizado en Beijing, China. Salinas-Rodriguez et. al~\cite{Salinas-Rodríguez_Ortiz-Pérez_García-Cuellar_Martínez-Salinas_2019} tomaron datos en México encontrando una correlación inversa significativa entre la precipitación y la concentración de $PM_{2.5}$. Takemoto et. al.~\cite{Takemoto_Medeiros_Ferreira_Mariani_2019} realizaron un estudio en la ciudad de Brasil, encontraron una correlación significativa entre la humedad relativa y la concentración de $PM_{2.5}$.

Teniendo esto en cuenta, se observa que la humedad relativa también puede influir en la concentración de $PM_{2.5}$ en el aire. En un estudio realizado en la ciudad de Delhi, India,~\cite{Praveen_2018}. Encontraron una correlación positiva entre la humedad relativa y la concentración de $PM_{2.5}$. De igual manera otro estudio realizado en la ciudad de Taipéi, Taiwán, por~\cite{Yang_2020}, encontró que la humedad relativa estaba inversamente relacionada con la concentración de $PM_{2.5}$. Estos resultados sugieren que los efectos de la humedad relativa en las concentraciones de $PM_{2.5}$ pueden variar según la región y las condiciones climáticas regionales.

A pesar de la importancia de estos factores para las concentraciones de $PM_{2.5}$ en el aire, se sabe relativamente poco sobre sus interacciones en ciudades de gran altitud como la ciudad de estudio. Por lo tanto, el objetivo de este estudio es correlacionar la concentración de $PM_{2.5}$ en el aire en función de la humedad relativa y precipitación en la ciudad del Cusco, ciudad que se ubica a una altitud de más de 3000 metros sobre el nivel del mar, con el fin de mejorar la comprensión y los factores que contribuyen a la contaminación del aire. Este análisis se realizará utilizando datos históricos sobre $PM_{2.5}$, precipitación y humedad relativa recopilados durante varios años por la estación High Vol del centro de energía y atmósfera de la Unsaac.

En este estudio, se investiga la relación entre los niveles de $PM_{2.5}$ y las condiciones meteorológicas, específicamente la precipitación y la humedad relativa. Se sabe que la precipitación y la humedad relativa pueden influir en la concentración de $PM_{2.5}$ en el aire al afectar la formación, la deposición y el transporte de partículas~\cite{Li_Li_Li_Li_Liu_Bai_2020}.

Estos resultados sugieren que los efectos de la humedad relativa en las concentraciones de $PM_{2.5}$ pueden variar según la región y las condiciones climáticas regionales. A pesar de la importancia de estos factores para las concentraciones de $PM_{2.5}$ en el aire, se sabe relativamente poco sobre sus interacciones en ciudades de gran altitud como la ciudad del Cusco.

La metodología ha consistido en analizar los datos obtenidos en el Centro de Investigación de Energía y Atmosfera - UNSAAC en la ciudad del Cusco que cuenta con una cabina móvil o Shelter equipada con analizdores de gases de efecto invernadero y un equipo muestreador de material particulado. Las mediciones se realizaron desde 2016 al 2018 con el objetivo de muestrear las concentraciones del material particulado ($PM_{2.5}$) ubicadas en diferentes sitios de monitoreo en la ciudad del Cusco, la humedad relativa y precipitación
se obtuvieron de datos proporcionados por el SENAHMI ubicado en la granja Kayra perteneciente a la UNSAAC. Asi se pudo establecer una correlación de concentración del material particulado ($PM_{2.5}$) y las variables meteorológicas (precipitación y humedad relativa). Los resultados de la metodologıa desarrollada nos permitió establecer una correlación de concentración del material particulado ($PM_{2.5}$) y las variables meteorológicas (precipitación y humedad relativa) para la cual se usaron los valores calculados por el muestreador de material particulado y aplicando Python para el tratamiento de datos, Python es un lenguaje de programación de propósito general que se caracteriza por su facilidad de uso y su capacidad para trabajar con diferentes tipos de datos y estructuras. Es ampliamente utilizado en la comunidad de programación debido a su sintaxis simple y legible, y a la gran cantidad de bibliotecas y módulos disponibles para su uso~\cite{van1995python}, para este estudio se opto por el uso de librerías de Python como Pandas que es una biblioteca diseñada para el análisis y manipulación de datos. Pandas es una herramienta esencial para el análisis de datos en Python. Ofrece una amplia gama de estructuras de datos, como Series y DataFrames, que son fáciles de manipular y analizar~\cite{mckinney2010data}. Conjuntamente con Pingoin una biblioteca de estadísticas de Python que proporciona una amplia variedad de pruebas estadísticas y herramientas para el análisis de datos. Pingouin ofrece una gama de funciones estadísticas útiles, y puede ser una alternativa a otras bibliotecas de estadísticas de Python, como SciPy~\cite{vallat2018pingouin}. Ya que este estudio tiene objetivo adecuar modelos de regresión lineal, tenemos en consideración el coeficiente de correlación de pearson, mencionado que este es una medida de dependencia que nos va permitir calcular el grado de covariación de dos variantes cuantitativas, sólo se aplica cuando las variables se encuentran relacionados de forma lineal, de la misma manera las variables deben de ser continuas~\cite{Dagnino_2014}. De igual manera se emplea el criterio analítico de Spearman que es un análisis no paramétrico que se usa para muestras con distribución normal dado que este tipo de análisis de correlación tiene el objetivo de medir la relación y asociación de una variable en referencia a otra con el fin de poder determinar si esta pueda servir como predictor; así mismo este coeficiente de correlación analiza también variables nominales y ordinales~\cite{barrera2014uso}.

La importancia de la investigación ha sido contribuir con el conocimiento sobre los diferentes factores meteorológicos que afectan en la contaminación del aire por $PM_{2.5}$ en la ciudad del Cusco. Conocer esta relación entre las diferentes variables puede ayudar a identificar patrones y factores que contribuyen en la contaminación del aire en la ciudad del Cusco, la información puede servir para adoptar medidas para disminuir la contaminación del aire.

\section{Datos}
Los datos considerados para el estudio son distintos meses del 2016 al 2018, estos son valores obtenidos en los muestreos de material particulado. La tabla \ref{table:promedio de pm2.5} muestra la concentración de $PM_{2.5}$, humedad relativa y precipitación.
\begin{table}[H]
	\centering
 \resizebox{.8\textwidth}{!} {
		\begin{tabular}{c c l c c c}
			\toprule
                N & Año & Mes & \multirow{2}{3.5cm}{\centering Promedio mensual $PM_{2.5}$ ($\frac{\mu g}{m^3}$)} & HR media(\%) & Precipitación ($mm$) \\
	       &&&&&\\		
   \midrule
			1 & 2016 & Noviembre & 108.11 & 56 &  28 \\
			2 & 2016 & Diciembre & 54.85 & 70 &  89.8 \\
			3 & 2017 & Mayo & 65.04 & 73 &  11.2 \\
			4 & 2017 & Julio & 152.44 & 63 & 0 \\
			5 & 2017 & Agosto & 121.38 & 60 &  8.4 \\
			6 & 2017 & Setiembre & 68.53 & 61 & 19 \\
			7 & 2017 & Octubre & 81.87 & 62 &  33.7 \\
            8 & 2017 & Diciembre & 75.99 & 68 &  101.7 \\
			9 & 2018 & Enero & 51.6 & 75 &  154.8 \\
			10 & 2018 & Febrero & 37.99 & 75 &  162.5 \\
			11 & 2018 & Marzo & 68.48 & 77 & 146.3 \\
			12 & 2018 & Abril & 81.62 & 72 &  20.3 \\
			13 & 2018 & Mayo & 72.85 & 68 & 0.2 \\
			14 & 2018 & Junio & 62.22 & 69 &  16 \\
                15 & 2018 & Agosto & 102.98 & 69 &  7.1 \\
			\bottomrule
	   \end{tabular}
    }
	\caption{Valores promedio de la concentración mensual de $PM_{2.5}$, humedad relativa y precipitación}
	\label{table:promedio de pm2.5}
\end{table}
\vspace{0.15cm}

\section{Metodología}

Se ha considerado datos correspondientes a la concentración mensual de material particulado 2.5 ($PM_{2.5}$) medido durante el periodo 2016-2018, las mediciones se realizó empleando un muestreador de material particulado de alto volumen por parte del Centro de Energía y Atmósfera (UNSAAC). Los datos de precipitación y humedad relativa se obtuvieron del SENAHMI (Kayra-UNSAAC).

Para estimar la relación estadística entre dos variables cuantitativas X y Y, se puede hacer uso de dos coeficiente de correlación (Pearson y Spearman)-técnicas utilizadas en la minería de datos-definidos por las siguientes expresiones \ref{equation:e1} y \ref{equation:e2} correspondientemente:

\begin{equation}
\label{equation:e1}
\ r_{p} = \frac{\sum\limits_{i=1}^{n}(X_i-\bar{X})(Y_i-\bar{Y})}{n\sigma_X\sigma_Y} \
\end{equation}

\begin{equation}
\label{equation:e2}
\ r_s=\frac{6\sum\limits_{i=1}^{n} d_i^{2}}{n(n^{2}-1)}\
\end{equation}\\

donde $X_i$ y $Y_i$ son datos i-ésimos, $\overline{X}$ y $\overline{Y}$ son los valores promedios y $\sigma_X$ y $\sigma_Y$ son los valores de la desviación estándar para los atributos $X$ y $Y$ respectivamente y $n$ es el número total de datos. Además $d_{i}^{2}$ es distancia entre los rangos. En general, valores de $r_{p}$ y $r_{s}$ cercanos a 1 y -1 corresponden a una correlación positiva o negativa, lo que significa que existe una dependencia relativa entre los dos tipos de datos $X$ y $Y$; la dependencia es relativa, ya que un valor de $r_{p} = 1$ y $r_{s} = 1$ (0-1) corresponden a una nula dispersión en los datos, y a su vez corresponden a un caso ideal. Sin embargo, valores de $r_{p} = 0$ y $r_{s} = 0$ (Tabla \ref{table:interpretacion correlacion}), corresponden a una nula relación entre variables $X$ y $Y$ o una muy grande dispersión de los datos. para establecer una interpretación del valor obtenido de $r_{p}$ y $r_{s}$ y ver si los datos $X$ y $Y$ están o no relacionados, y dicha relación no es producto de la aleatoriedad, se calcula el valor de la distribución $t$ de Student mediante la expresión:

\begin{equation}
\label{equation:e3}
t = r \frac{\sqrt{n - 2}}{\sqrt{1 - r^2}}
\end{equation}

donde $r$ es el coeficiente de correlación de Pearson o Spearman y  $n$ es el tamaño de la muestra.\\

\begin{table}[H]
	\centering
 \resizebox{.9\textwidth}{!} {
		\begin{tabular}{c c c c c c}
			\toprule
                \multicolumn{1}{l}{} & Muy alta & Alta & Media & Baja & Muy baja \\
			\midrule
			Positiva &  $1 \leq r < 0.80$ & $+0.80 \leq +0.60$ & $+0.60 \leq +0.40$ &  $+0.40 \leq +0.20$ & $+0.20 \leq 0$ \\
			Negativa &  $-1 \leq r < -0.80$ & $-0.80 \leq -0.60$ & $-0.60 \leq -0.40$ &  $-0.40 \leq -0.20$ & $-0.20 \leq 0$ \\
			\bottomrule
	   \end{tabular}
    }
	\caption{Interpretación de los valores de correlación r de Pearson y $\rho$ de Spearman}
 
 Fuente: Libro Estadística para no estadísticos ~\cite{G_Domínguez}
	\label{table:interpretacion correlacion}
\end{table}

El valor obtenido de $t$ se compara con el valor obtenido de la tabla $t$ de Student para un cierto nivel de significancia (se ha establecido ${\alpha}=0.05$-correspondiente a un resultado que normalmente es aceptable). Por otro lado, según la tabla $t$ de Student para $n=15$, $t=1.7613$. Además, para este estudio se ha establecido una hipótesis nula ($h_{0}$: no existe correlación significativa entre entre el par de variables) y una hipótesis alterna ($h_{a}$: existe correlación significativa entre el par de variables). Los criterios que se utilizan para aceptar o rechazar la hipótesis nula son los siguientes: a) Si el valor de $t$ de Student obtenido de la tabla es menor que la calculada, entonces se dice que las variables están relacionadas; la correlación no proviene del azar(Se rechaza la hipótesis nula). b) Por el contrario, si la $t$ obtenida de las tablas es mayor que la calculada, entonces se dice que no hay correlación entre las variables, aunque el valor de $r_{p}$ o $r_{s}$ sean diferentes de $0$; es decir, se acepta la hipótesis nula.\\

Para calcular el p-valor a partir del valor estadístico t, primero se necesita conocer los grados de libertad (gl) del estadístico t y luego usar una tabla de distribución t-Student o una calculadora estadística que te proporcione el valor de probabilidad. La fórmula general para calcular el p-valor a partir del valor de t y los grados de libertad es:

\begin{equation}
\label{equation: e4}
p-valor = P(T > t),  t=positivo
\end{equation}
\begin{equation}
\label{equation: e5}
p-valor = P(T < t),  t=negativo
\end{equation}

Donde $T$ es la distribución $t-Student$ con $gl$ grados de libertad.

Es importante recordar que el p-valor representa la probabilidad de obtener un valor de $t$ igual o más extremo que el observado, bajo la hipótesis nula. Si el p-valor es menor que el nivel de significancia establecido, se rechaza la hipótesis nula y se concluye que existe evidencia estadística para aceptar la hipótesis alternativa( hay evidencia significativa que existe correlación).

Para la estadística, en este estudio utilizaremos las librerías Pandas, StatsModels y Pingouin de Python, que proporcionan funciones para calcular el valor de $p-valor$ directamente a partir del coeficiente de correlación de Pearson o Spearman y el tamaño de la muestra.\\

\begin{figure}[H]
	\centering
     \includegraphics[width=\textwidth]{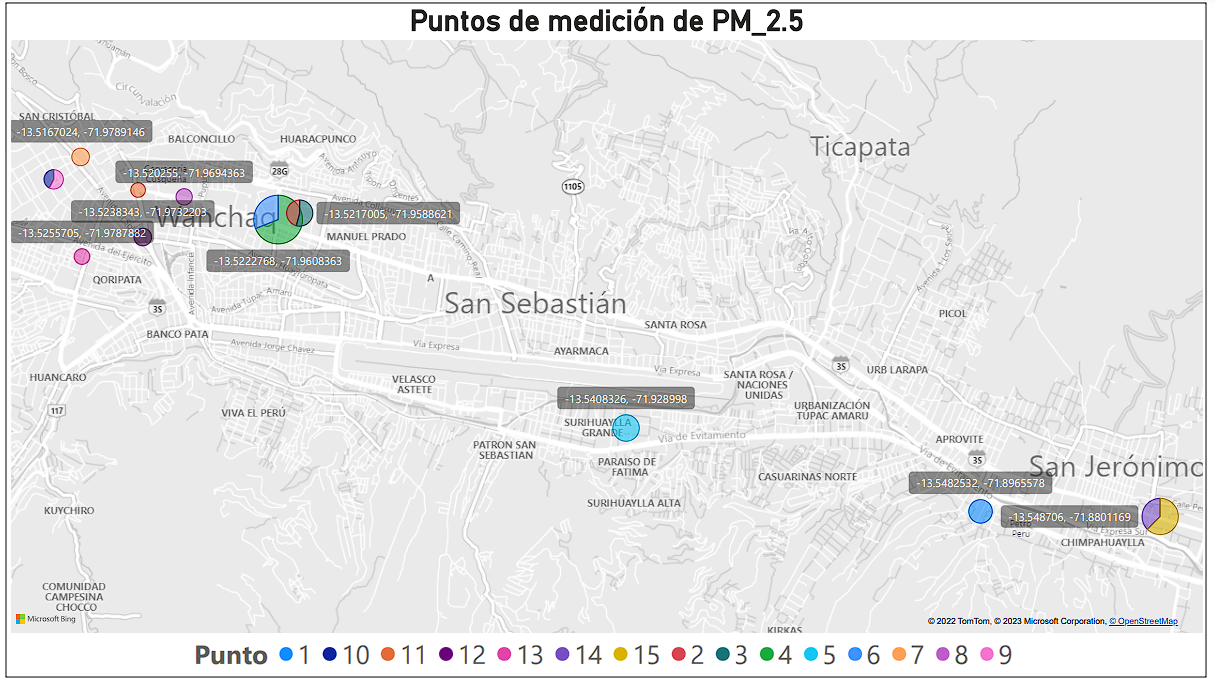}
	\caption{ Mapa de puntos de medición de $PM_{2.5}$ en función de la tabla \ref{table:pm2.5}. }	
    \label{img:mapa 1}
    
    \centerline{Fuente: Elaboración propia }
\end{figure}

En primer lugar,para el cálculo del coeficiente de correlación de Pearson y Spearman, las variables $X$ y $Y$ corresponden a las mediciones que se hacen sobre la concentración de $PM_{2.5}$ y la precipitación. En segundo lugar, las variables $X$ y $Y$ corresponden a las mediciones de concentración de $PM_{2.5}$ y Humedad Relativa. Ambos medidos durante el periodo 2016-2018. El estudio se realizó con los datos obtenidos de 15 puntos diferentes (tabla \ref{table:pm2.5}), ubicados y distribuidos geográficamente en la parte urbana de la ciudad del Cusco (mapa \ref{img:mapa 1}).


Los 15 puntos de medición están ubicados a distancias considerables y se encuentran distribuidas en diferentes distritos de la ciudad (tabla \ref{table:pm2.5}). Por cada mes, se tiene 1 dato promedio de $PM_{2.5}$, precipitación y húmeda relativa; los equipos de medición en algunos puntos arrojaron datos que son erróneos (números negativos o vacíos), por lo que se eliminaron mediante un proceso de limpieza de datos.

\begin{table}[H]
    \centering
    \resizebox{.9\textwidth}{!} {
    \begin{tabular}{lr}
    \toprule
        \multirow{2}{10cm}{Lugar de Medida}     & \multirow{2}{4cm}{Promedio mensual\\de $PM_{2.5}$($\mu g/m^3$)} \\
        & \\
    \toprule
    Punto 1 - San Jerónimo - Ingreso a ladrilleras              & $108.11$ \\
    Punto 2 - UNSAAC - Pabellón de Educación              &  $54.85$ \\ 
    Punto 2 - UNSAAC - Pabellón de Educación              &  $65.04$ \\  
    Punto 3 - UNSAAC - MULTIRED              & $152.44$ \\ 
    Punto 4 - San Sebastián - APV Paraíso de Fátima              & $121.38$ \\
    Punto 3 - UNSAAC - MULTIRED              &  $68.53$ \\ 
    Punto 5 - Plaza de Armas de Cusco              &  $81.87$ \\
    \multirow{2}{10cm}{Punto 6 - Colegio Clorinda Matto de Turner\\Punto 7 - Municipalidad de Wanchaq} &   \multirow{2}{.9cm}{$75.99$} \\  
    & \\ 
     \multirow{2}{10cm}{Punto 8 - Plaza San Francisco\\Punto 9 - Centro de Salud de Wanchaq}    &   \multirow{2}{.8cm}{$51.6$}  \\  
    & \\ 
    Punto 8 - Plaza San Francisco              &  $37.99$ \\   
    Punto 10 - Plaza Limacpampa           &  $68.48$ \\    
     \multirow{2}{10cm}{Punto 11 - Plaza Pumaqchupan\\Punto 12 - Calle Matará}  &   \multirow{2}{.9cm}{$81.62$} \\ 
     & \\ \
    Punto 13 - Centro de Salud de Belenpampa             &  $72.85$ \\  
    Punto 14 - Centro de Salud de San Jerónimo             &  $62.22$ \\ 
    Punto 14 - Centro de Salud de San Jerónimo             & $102.98$ \\  
    \bottomrule
    \end{tabular}
    }
    \caption{Concentración de $PM_{2.5}$, humedad Relativa y precipitación}
    \label{table:pm2.5}
\end{table}

En la primera fase del estudio se hizo gráficos de dispersión, para cada par de variables ($PM_{2.5}$-precipitación y $PM_{2.5}$-humedad relativa), con el fin de visualizar a priori la relación entre las variables. Así mismo, se realizó gráficas temporales de las mismas. Enseguida se realizó un mapa de calor de coeficientes de correlación de Pearson y Spearman, en esta se puede observar la relación entre las variables correspondientes. Finalmente realizamos un cuadro de correlación para el par $PM_{2.5}$-precipitación y $PM_{2.5}$-humedad relativa. En este cuadro se aprecia el $p-valor$ obtenido y el poder de los datos para cada tipo de correlación.

\section{Resultados y Discusión}

\begin{figure}[H]
	\centering
     \includegraphics[width=.7\textwidth]{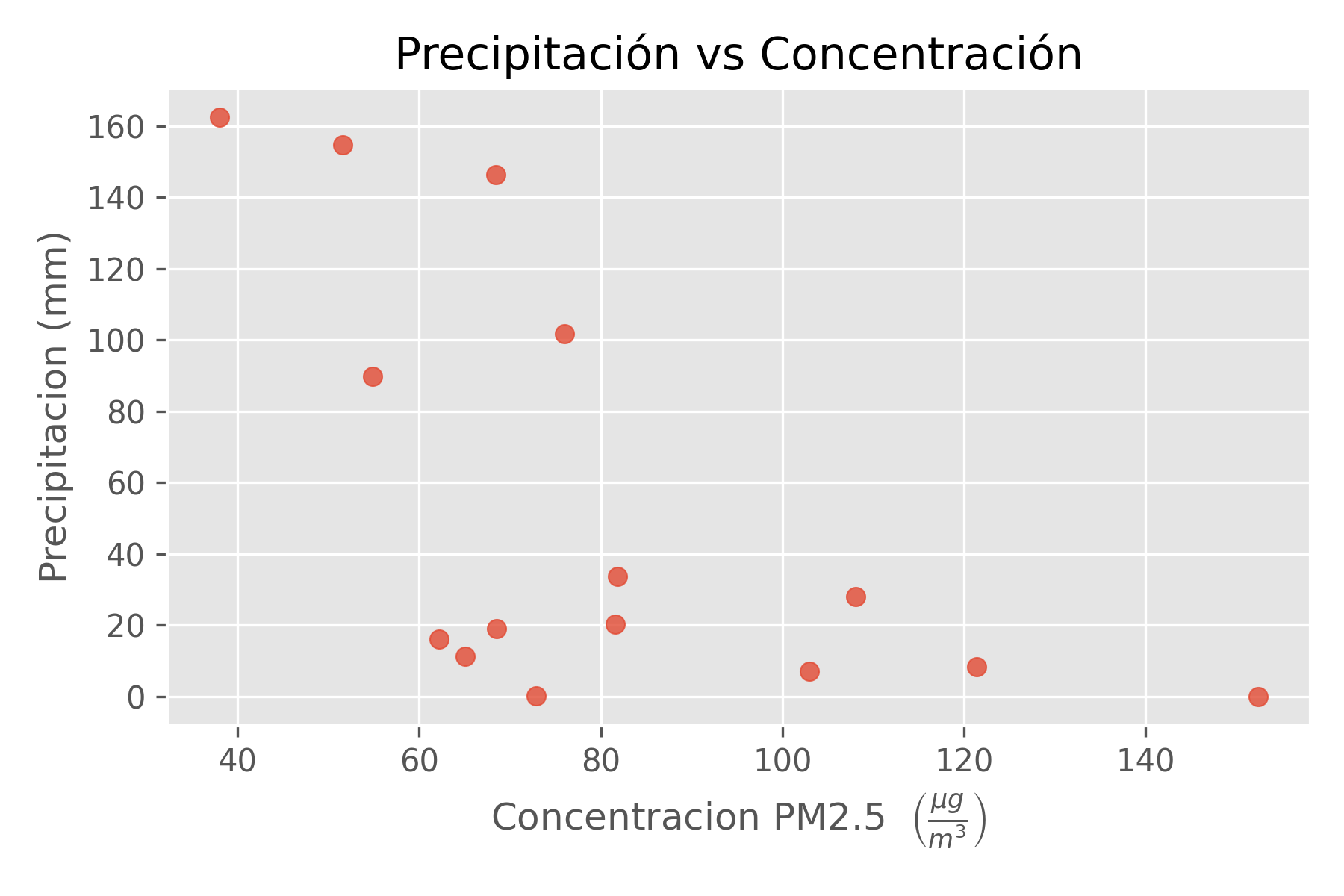}
	\caption{Gráfica de dispersión de concentración de $PM_{2.5}$ vs precipitación}	
    \label{img:scatter-con-prec}
    
    \centerline{Fuente: Elaboración propia }
\end{figure}

\begin{figure}[H]
	\centering
     \includegraphics[width=.7\textwidth]{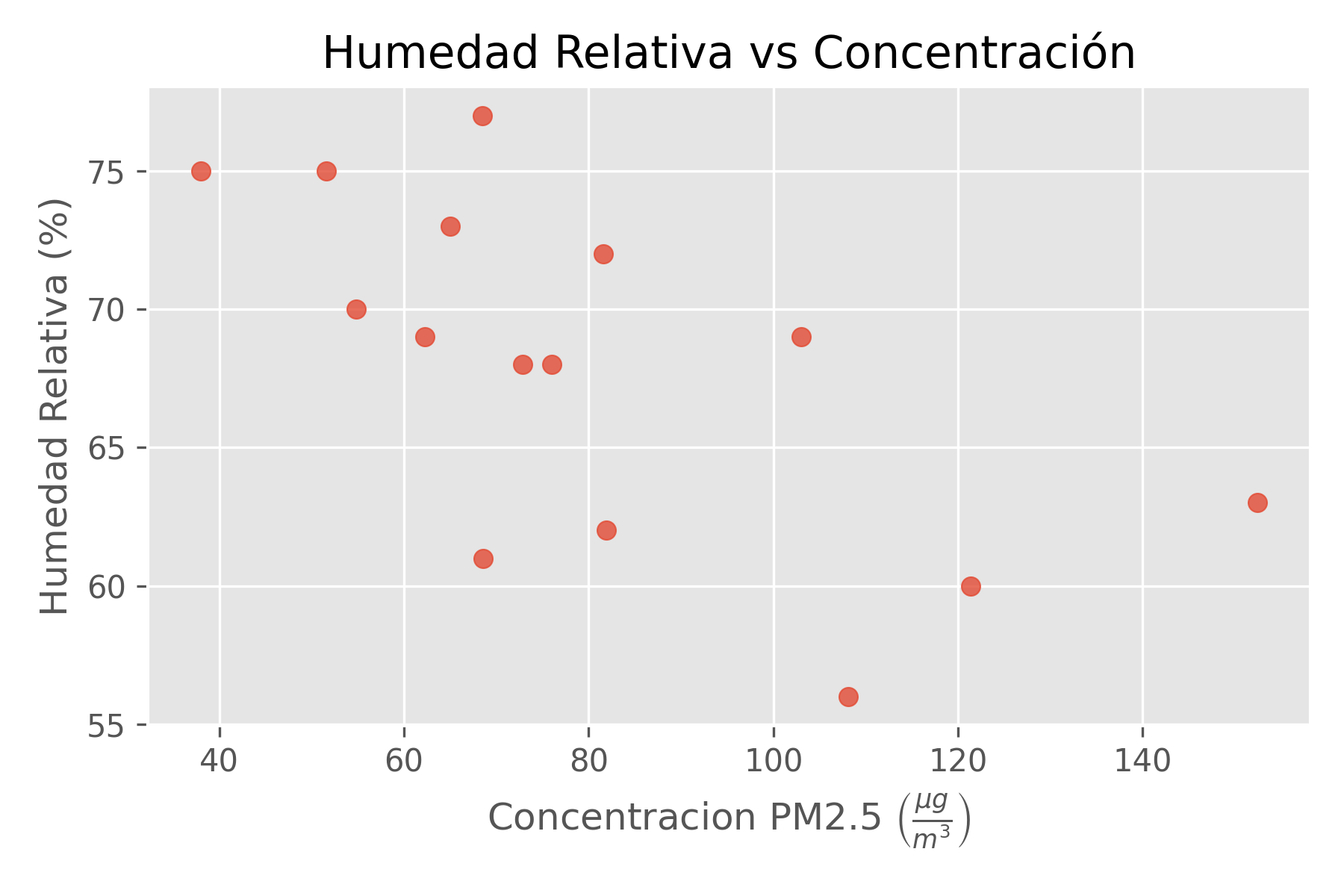}
	\caption{Gráfica de dispersión de concentración vs humedad relativa}	
    \label{img:scatter-con-hum}
    
    \centerline{Fuente: Elaboración propia }
\end{figure}

La gráfica de dispersión (figura \ref{img:scatter-con-prec}) muestra la relación entre los datos de concentración de $PM_{2.5}$ y la precipitación. En esta, se puede notar que a medida que la precipitación sufre una disminución, la concentración de $PM_{2.5}$ sufre un aumento; es decir, una relación inversa entre el par de variables. Similarmente, se logra apreciar en la figura \ref{img:scatter-con-hum} para los datos de humedad relativa y concentración de $PM_{2.5}$. Esto se podrá constatar más adelante cuando se haga el análisis estadístico para cada caso.\\

\begin{figure}[H]
	\centering
     \includegraphics[width=.85\textwidth]{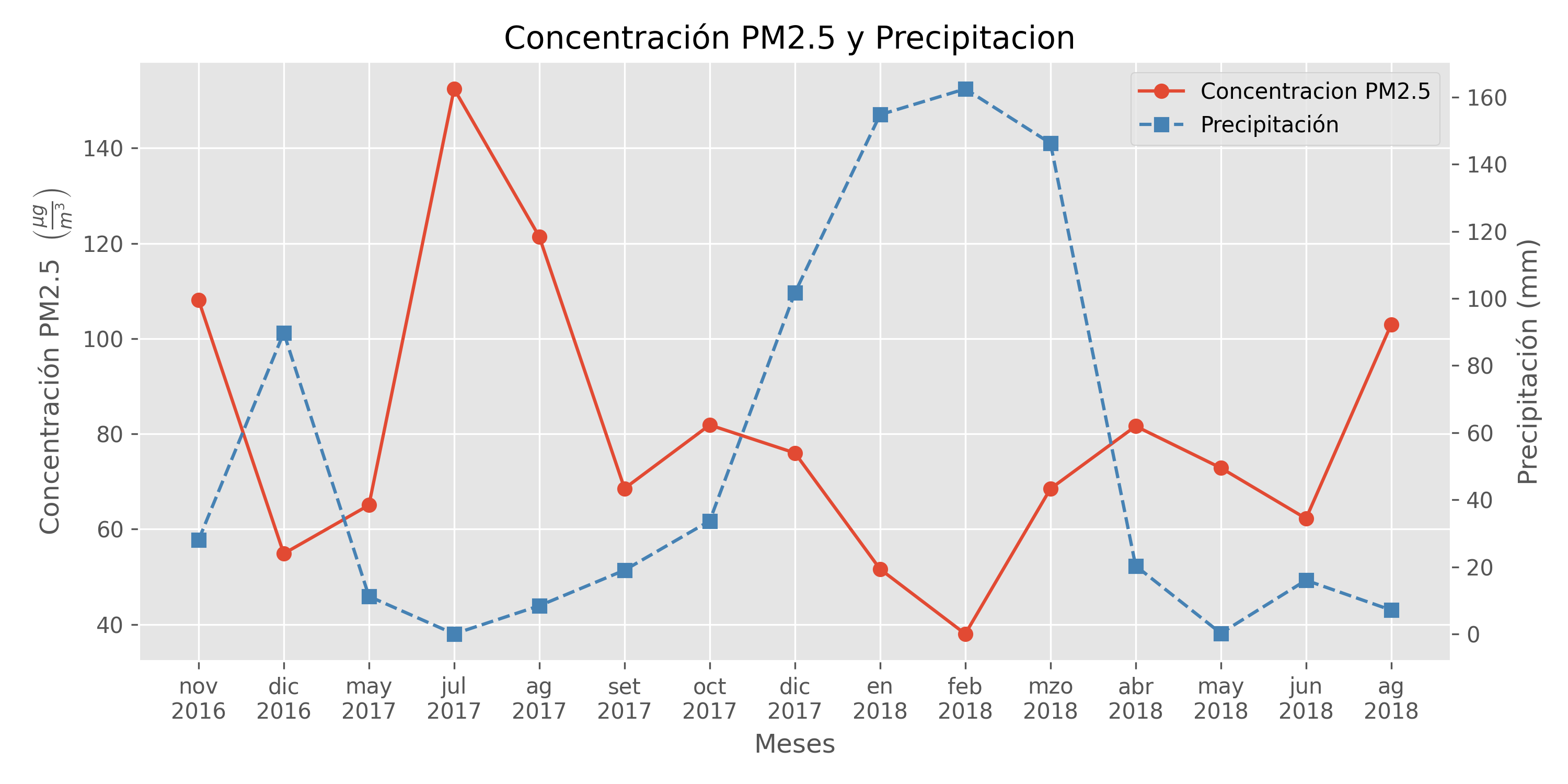}
	\caption{Gráfica de evolución temporal de la concentración de PM2.5 y precipitación para el periodo nov-2016 hasta ag-2018}	
    \label{img:plot-con-prec}
    
    \centerline{Fuente: Elaboración propia }
\end{figure}

Otra forma de visualizar la correlación entre las variables es mediante la figura \ref{img:plot-con-prec} que muestra la evolución temporal de la concentración de $PM_{2.5}$ en función de la precipitación. Esta figura muestra la existencia de correlación negativa entre ambas variables. Según la tabla \ref{table:promedio de pm2.5}, en la figura  \ref{img:plot-con-prec} se observa que la concentración de $PM_{2.5}$ disminuye de noviembre($108.11\ \mu g/m^3$) a diciembre($54.85\ \mu g/m^3$) en el año 2016. Al mismo tiempo, si observamos la precipitación aumenta, en ese mismo periodo de tiempo, de noviembre($28\ mm$) a diciembre($89.8\ mm$). Esta disminución y aumento, respectivamente, cobran sentido al tratarse de las diversas temporadas de precipitación de la ciudad Cusco, debido a sus estaciones muy marcadas.

\newpage

Por otro lado, de mayo ($65.04\ \mu g/m^3$) a julio ($152.44\ \mu g/m^3$) del 2017 se observa un aumento de la concentración de $PM_{2.5}$. Así mismo, si observamos la precipitación de mayo ($11.2\ mm$) a julio ($0\ mm$) disminuye lo cual se debe a que en el mes de julio no se presenta lluvias en la ciudad de Cusco . Seguidamente, al observar la concentración de $PM_{2.5}$ de agosto ($121.38\ \mu g/m^3$) del 2017 a febrero ($37.99\ \mu g/m^3$) del 2018 notamos que la concentración de $PM_{2.5}$ disminuye, esto se debe al aumento de precipitación en estos meses: agosto ($8.4\ mm$) del 2017 a febrero ($162.5\ mm$) del 2018.

Finalmente, se pone en manifiesto el comportamiento de la concentración de $PM_{2.5}$ de marzo ($121.38\ \mu g/m^3$) al mes de agosto ($37.99\ \mu g/m^3$) del 2018 se observa un aumento. Ello es debido a la disminución de lluvias(precipitación) en estos meses: marzo ($146.3\ mm$) a agosto ($7.1\ mm$) del 2018.

\begin{figure}[H]
	\centering
     \includegraphics[width=.85\textwidth]{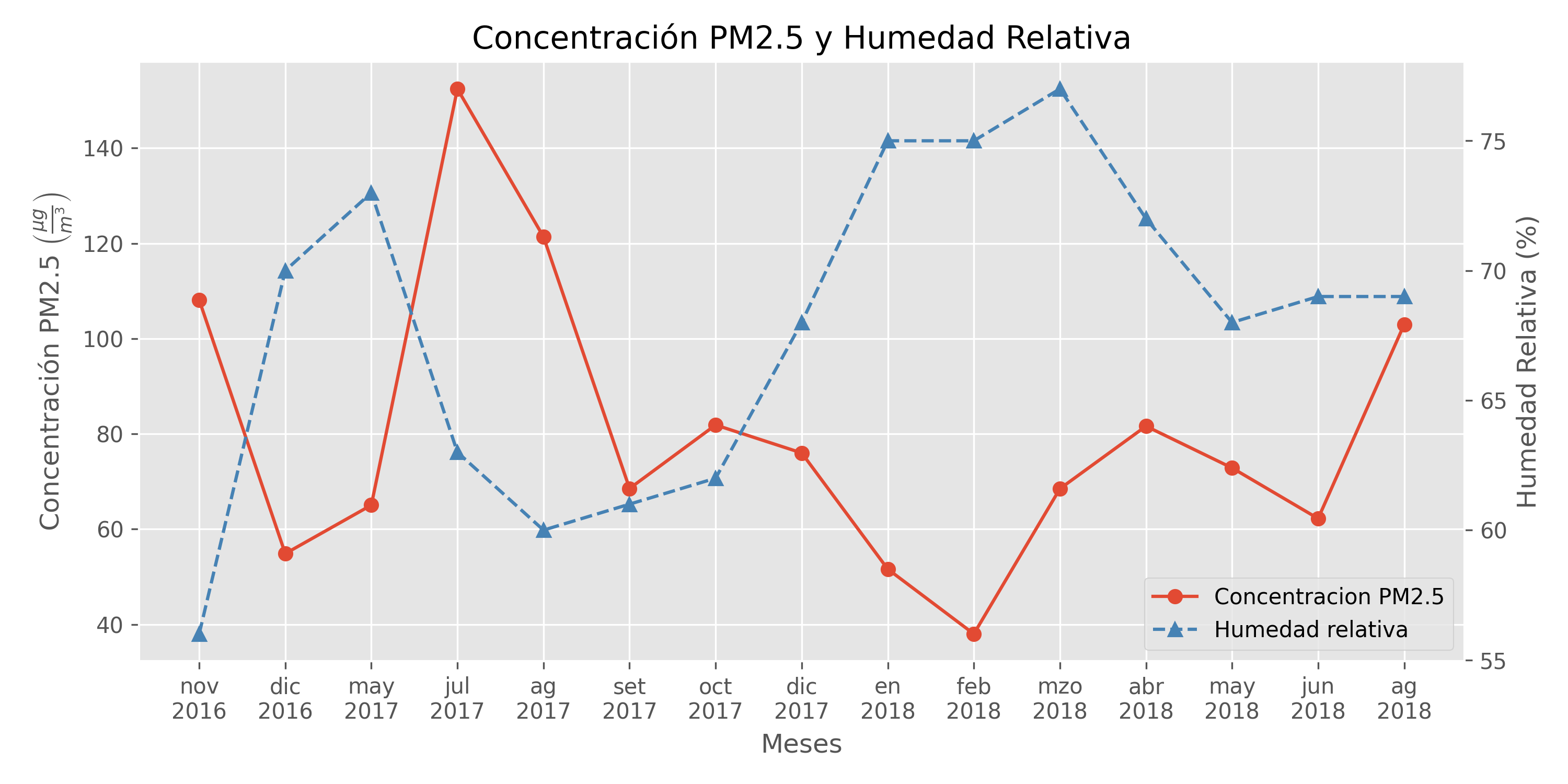}
	\caption{Gráfica de evolución temporal de la concentración de PM2.5 y humedad relativa para el periodo nov-2016 hasta ag-2018}	
    \label{img:plot-con-hum}
    
    \centerline{Fuente: Elaboración propia }
\end{figure}

De forma similiar a la anterior, en la gráfica  \ref{img:plot-con-hum} se muestra la evolución temporal del par $PM_{2.5}$-humedad relativa. En ese sentido, se aprecia que la concentración de $PM_{2.5}$ disminuye de noviembre ($108.11\ \mu g/m^3$) a diciembre ($54.85\ \mu g/m^3$) en el año 2016. A su vez, al observar la la humedad relativa se nota un aumento de noviembre ($56\ \%$) a diciembre ($70\ \%$).

Por otro lado, de mayo ($65.04\ \mu g/m^3$) a julio ($152.44\ \mu g/m^3$) del 2017 se observa un aumento de la concentración de $PM_{2.5}$. Así mismo, se observa la precipitación de mayo ($73\ \%$) a julio ($63\ \%$) disminuye. Seguidamente, al observar la concentración de $PM_{2.5}$ de agosto ($121.38\ \mu g/m^3$) del 2017 a febrero ($37.99\ \mu g/m^3$) del 2018 se pone en manifiesto que la concentración de $PM_{2.5}$ disminuye, esto se debe al aumento de precipitación en estos meses: agosto ($60\ \%$) del 2017 a febrero ($75\ \%$) del 2018.

Finalmente, se pone en manifiesto el comportamiento de la concentración de $PM_{2.5}$ de marzo ($121.38\ \mu g/m^3$) al mes de agosto ($37.99\ \mu g/m^3$) del 2018 se observa un aumento. Ello se debe a la disminución de la humedad relativa en estos meses: marzo ($77\ \%$) a agosto ($69\ \%$) del 2018.

\begin{figure}[H]
	\centering
     \includegraphics[width=.5\textwidth]{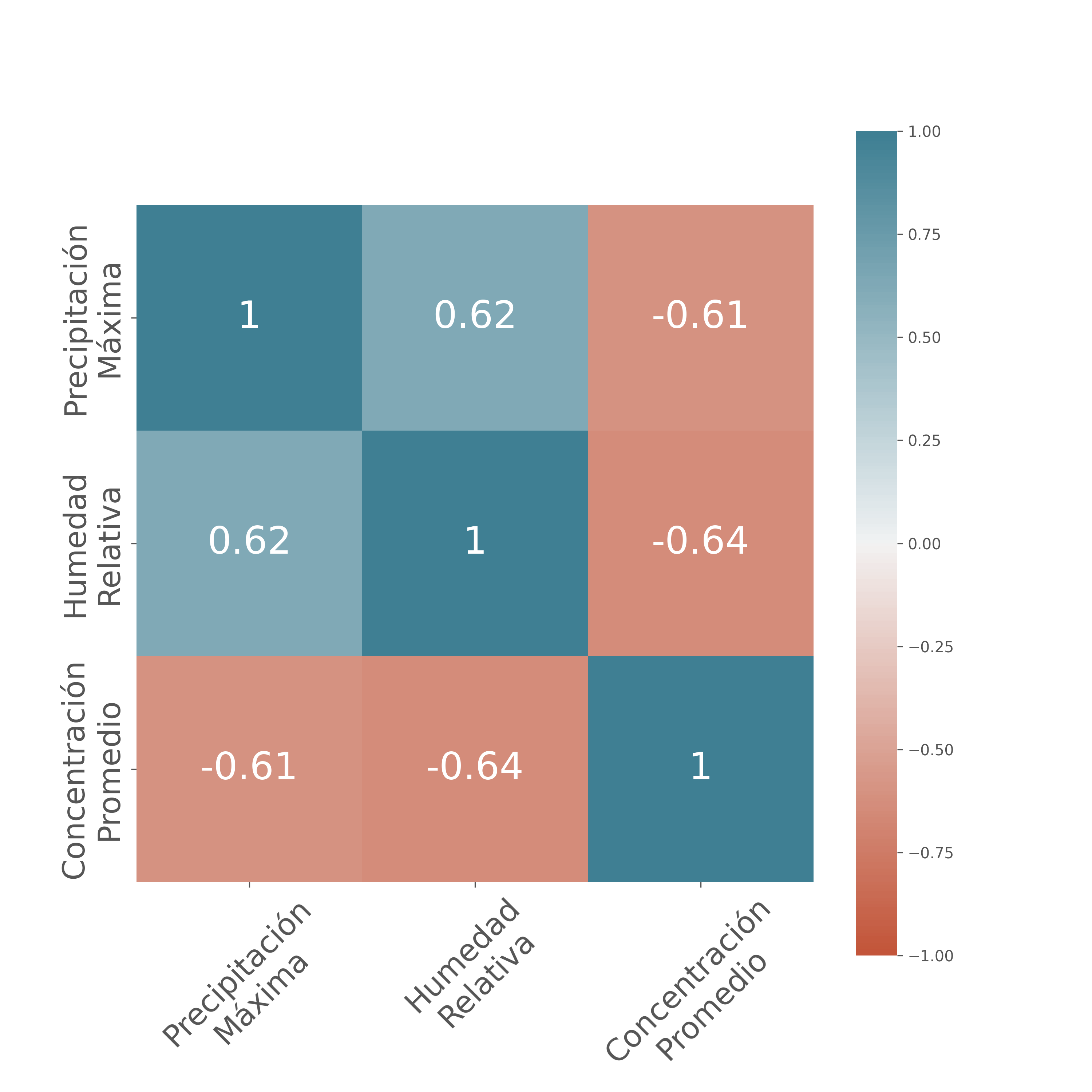}
	\caption{Mapa de calor de coeficientes de correlación de Pearson para la concentración de $PM_{2.5}$, humedad y precipitación}	
    \label{img:matrix-corr-pearson}
    
    \centerline{Fuente: Elaboración propia }
\end{figure}

La figura \ref{img:matrix-corr-pearson}-mapa de calor de coeficientes de correlación de Pearson para las variables $PM_{2.5}$, precipitación y humedad relativa. En esta, podemos apreciar el valor obtenido del coeficiente de correlación para cada par de variables. En primer lugar, se tiene un valor $r_{s}=-0.61$ para el par $PM_{2.5}$-precipitación. En segundo lugar, se tiene un valor $r_{s}=-0.64$ para el par $PM_{2.5}$-humedad relativa. Es evidente la correlación negativa en ambos casos. Al usar Pearson tenemos que tener en cuenta que la relación de datos que se busca es lineal, además se busca la causalidad del par de variables, sin embargo, no siempre se puede afirmar ello.

\begin{figure}[H]
	\centering
     \includegraphics[width=.5\textwidth]{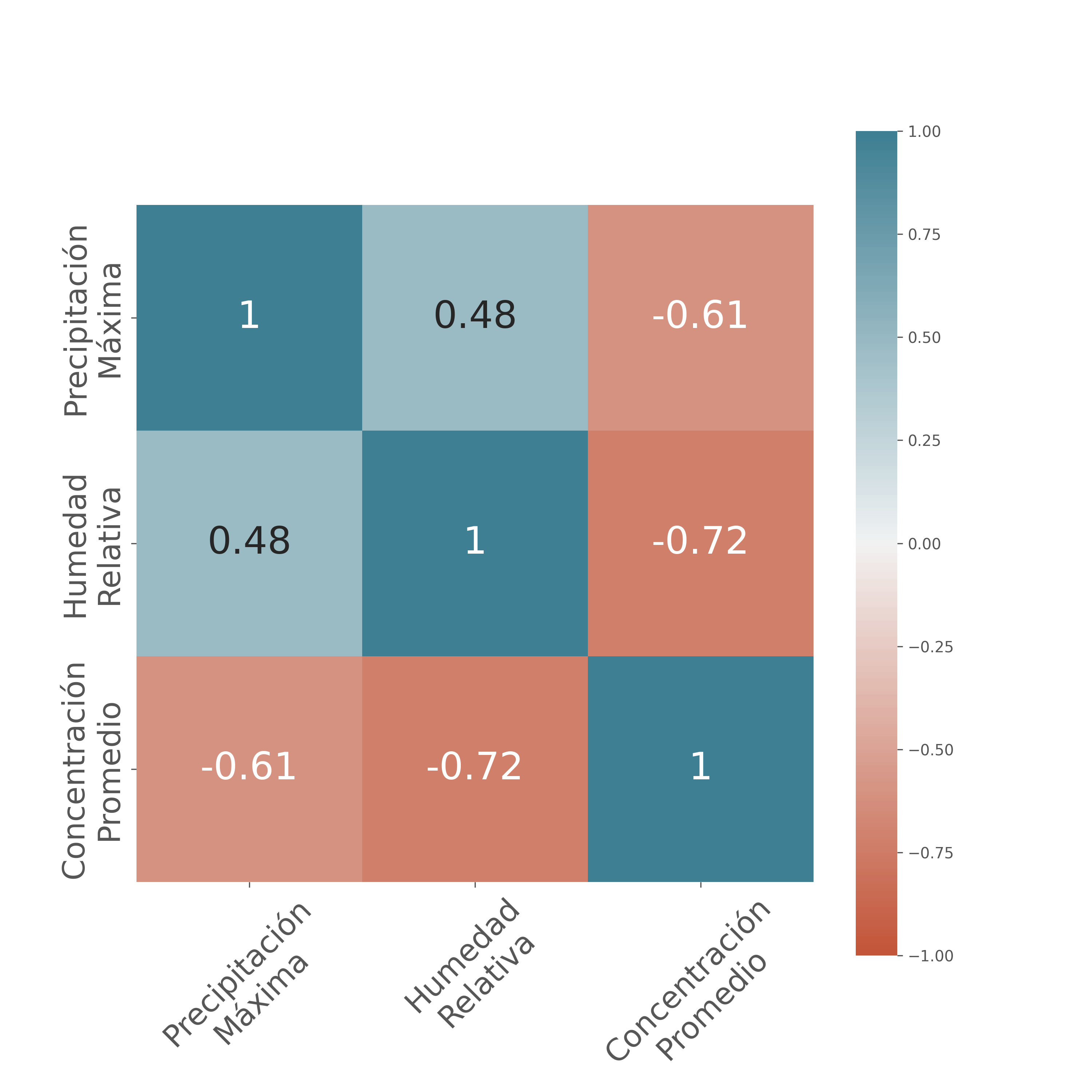}
	\caption{Mapa de calor de coeficientes de correlación de Spearman para la concentración de $PM_{2.5}$, humedad y precipitación}	
    \label{img:matrix-corr-spearman}
    
    \centerline{Fuente: Elaboración propia }
\end{figure}

De igual forma, la figura \ref{img:matrix-corr-spearman}-mapa de calor de coeficientes de correlación de Spearman para las variables $PM_{2.5}$, precipitación y humedad relativa. En ella es posible visualizar el valor obtenido del coeficiente de correlación para cada par de variables. En primer lugar, se tiene un valor $r_{p}=-0.61$ para el par $PM_{2.5}$-precipitación y un valor $r_{s}=-0.72$ para el par $PM_{2.5}$-humedad relativa. Es evidente la correlación negativa para cada par de variables. Además, al usar Spearman tenemos que tener en cuenta que no se busca la causalidad de variables.

Finalmente se tienen dos tablas de comparación de los coeficientes de correlación de Pearson y Spearman. Las tablas \ref{table:pm2-5-humedad-corr-matrix} y \ref{table:pm2-5-precipitacion-corr-matrix} muestran correlaciones entre cada par de variables($PM_{2.5}$-humedad relativa y $PM_{2.5}$-precipitación). Se utiliza el paquete Pingouin de Python para calcular diferentes tipos de correlaciones. En este caso, se están calculando dos tipos diferentes de correlación entre cada par de variables. La correlación de Pearson mide la relación lineal entre dos variables, la correlación de Spearman mide la relación monótona (creciente o decreciente) entre dos variables.

Los resultados muestran en una tabla, con el tamaño  de muestra $n$, el coeficiente de correlación de Pearson $r_{p}$ y Spearman $r_{s}$. Así mismo, se puede observar los intervalos de confianza del $95\%$, así como también el $p-valor$ obtenido. Adicionalmente, se muestra la potencia estadística de las variables. 

La potencia estadística indica la probabilidad de que un análisis estadístico encuentre una diferencia o una relación real entre dos variables cuando esta realmente existe. La potencia estadística está influenciada por varios factores, como el tamaño de muestra, el nivel de significancia. En general, se considera que una potencia estadística del $80\%$ o más es suficiente para poder detectar efectos importantes en los datos.


\begin{table}[H]
    \centering
    \resizebox{.8\textwidth}{!} {
    \begin{tabular}{lccccc}
    \hline
                 & n  & r           &  CI$95 \%$      & p-val      & Potencia estadística \\
    \hline
        Pearson  & 15 & $-0.644521$ & $[-0.87, -0.2]$ & $0.009492 < \alpha = 0.05$ & $0.777794$           \\
        Spearman & 15 & $-0.723369$ & $[-0.9, -0.34]$ & $0.002304 < \alpha = 0.05$ & $0.901474$           \\
    \hline
    \end{tabular}
    }
    \caption{Cuadro de correlaciones de la concentración de $PM_{2.5}$ y la humedad relativa}
    \label{table:pm2-5-humedad-corr-matrix}
    \centerline{Fuente: Elaboración propia }

\end{table}

\begin{table}[H]
    \centering
    \resizebox{.8\textwidth}{!} {
    \begin{tabular}{lccccc}
    \hline
                 & n  & r           &  CI$95 \%$      & p-val      & Potencia estadística \\
    \hline
        Pearson  & 15 & $-0.60535$ & $[-0.85, -0.13]$ & $0.016789 < \alpha = 0.05$ & $0.704405$           \\
        Spearman & 15 & $-0.607143$ & $[-0.85, -0.14]$ & $0.016381 < \alpha = 0.05$ & $0.707864$           \\
    \hline
    \end{tabular}
    }
    \caption{Cuadro de correlaciones de la concentración de $PM_{2.5}$ y la precipitación máxima}
    \label{table:pm2-5-precipitacion-corr-matrix}
    \centerline{Fuente: Elaboración propia }

\end{table}

Debido a que $p-valor < \alpha = 0.05$, en ambas tablas [\ref{table:pm2-5-humedad-corr-matrix}][\ref{table:pm2-5-precipitacion-corr-matrix}], se acepta la hipótesis alterna y se rechaza la hipótesis nula; es decir, existe una correlación significativa entre el par de variables $PM_{2.5}$-precipitación y $PM_{2.5}$-humedad relativa. Esto se evidencia tanto para el coeficiente de correlación de Pearson y Spearman. Así mismo, para el par $PM_{2.5}$-humedad relativa, en la tabla \ref{table:pm2-5-humedad-corr-matrix} se observa una potencia estadística del $77.77\%$ y $90.1\%$  para Pearson y Spearman respectivamente.
Al mismo tiempo, para el par $PM_{2.5}$-precipitación, en la tabla \ref{table:pm2-5-precipitacion-corr-matrix} se observa una potencia estadística del $70.4\%$ para Pearson y $70.7\%$ para Spearman.

 

\section{Conclusiones}
 Los resultados del análisis de correlación utilizando los coeficientes de Pearson y Spearman indicaron una alta correlación negativa entre $PM_{2.5}$ y la humedad relativa y a su vez también con la precipitación, siendo estas que nos brindaron valores de correlación de Pearson [Figura \ref{img:matrix-corr-pearson}] entre la concentración promedio de material particulado $PM_{2.5}$ y la precipitación máxima con un valor de $-0.61$, la concentración promedio de material particulado $PM_{2.5}$ y la humedad relativa con un valor de $-0.64$, dándonos a entender que en función a la correlación de Pearson la correlación es negativa alta entre estas variables, según la tabla \ref{table:interpretacion correlacion}, así mismo los valores de correlación de Spearman [Figura \ref{img:matrix-corr-spearman}] entre la concentración promedio de material particulado $PM_{2.5}$ y la precipitación máxima nos dan un valor de $-0.61$, la concentración promedio de material particulado $PM_{2.5}$ y la humedad relativa nos dan un valor de $-0.72$, dándonos a entender que en función a la correlación de Spearman la correlación es negativa alta entre estas variables, según la Tabla \ref{table:interpretacion correlacion}.

Estos hallazgos sugieren que la humedad relativa y la precipitación tienen un impacto significativo en los niveles de $PM_{2.5}$ en la ciudad del Cusco. Por lo tanto, se pueden considerar medidas específicas para abordar los efectos de estos factores en la calidad del aire y reducir los niveles de material particulado $PM_{2.5}$. Además, los resultados del estudio pueden contribuir a futuras investigaciones sobre la relación entre algunas variables meteorológicas y la calidad del aire, y pueden ser relevantes para la formulación de políticas públicas y estrategias de mitigación en la ciudad del Cusco y a su vez normativas generales a nivel de nación.

\section*{Agradecimientos}
A la químico Amanda Olarte, por su apoyo en brindarnos conocimientos referentes a la físico quimica de la contaminación ambiental, al Magíster Marco A. Zamalloa por su apoyo en la búsqueda de herramientas de obtención de datos y trabajo de estas mismas, al Magíster Ricardo F. Camargo por su apoyo y motivación para publicar, al bachiller Cuiro, E. Weyner por su contribución con bibliografía relacionada a estadística.   \\
A la Universidad Nacional de San Antonio Abad del Cusco, que en conjunto al centro de Energía y Atmósfera, por ayudarnos con el uso de equipos y data sobre material particulado $PM_{2.5}$.

\printbibliography
\end{document}